\documentclass[12pt]{article}
\usepackage{graphicx,amsmath}
\usepackage{amssymb,amsthm} 
\usepackage{subfig}
\usepackage{array}
\usepackage{multirow}
\title{Deliverable navigation for multicriteria step and shoot IMRT treatment planning}
\author{David Craft and Christian Richter \\ \\ Department of Radiation Oncology \\ Massachusetts General Hospital}
\date{November 1, 2012} 
\begin{document}
\maketitle
\thispagestyle{empty}

\begin{abstract}
We consider Pareto surface based multi-criteria optimization for step and shoot IMRT planning.
By analyzing two navigation algorithms, we show both theoretically and in practice that the number
of plans needed to form convex combinations of plans during navigation can be kept small (much less than the theoretical maximum number needed in general, which is equal to the number of objectives for on-surface Pareto navigation). 
Therefore a workable approach for directly deliverable navigation in this setting is to segment 
the underlying Pareto surface plans and then enforce the mild restriction that only a small number of these plans are active at any time during plan navigation, thus limiting the total number of segments used in the final plan.
\end{abstract}

\section{Introduction}

Although intensity modulated radiation therapy (IMRT) has become a clinical standard for treating many cancers, including prostate, head and neck, pancreas, brain, and lung, the planning process -- which relies heavily on automated computer optimization -- remains a time consuming and iterative task. This is due to the fact that IMRT planning is inherently a search for a plan which balances conflicting desires, namely: prescribed dose to the target(s) and minimal dose to the organs at risk (OARs).  Multi-criteria optimization (MCO) has been introduced as a technique to allow the planner to explore the patient-specific tradeoffs, and has been shown to reduce treatment planning time and achieve superior plans for selected clinical scenarios \cite{Craft2011, hong}.

Under the general title of MCO fall two distinct approaches to treatment planning: prioritized optimization, also referred to as lexicographic optimization, and Pareto surface based MCO. In prioritized optimization, a sequence of optimizations are run, tackling the most important objectives first and then moving down the list of lower objectives \cite{breedveld, jee, wilkens, falkinger}. For each optimization run, the results of the previous higher priority optimizations are used as constraints to ensure that the higher priorities are respected.  This approach results in a single plan and if all goes well, this plan is used as the clinical treatment plan.  In Pareto surface based MCO the planner interacts with the planning system to explore dosimetric tradeoffs -- different Pareto optimal plans -- in real time.  For a given set of objectives and constraints, an IMRT plan is Pareto optimal if it is feasible (satisfies the constraints) and it is not possible to improve on any objective without worsening at least one other objective. The full collection of Pareto optimal plans, typically an infinite set, is referred to as the Pareto surface. A practical technique for exploring this infinite set involves precomputing a modest number ($<$50) of Pareto optimal plans scattered across the Pareto surface and approximating the continuous Pareto surface by forming convex combinations of these base plans in response to user requests. For example, by way of a set of graphical user interface slider bars, a user might make a request to reduce spinal cord dose, at which point the system computes a convex combination of the base plans which has an improved spinal cord dose. Computing these convex combinations can be done fast enough to give the user the experience of continuous real time navigation.

Exploring a Pareto surface of treatment plan possibilities has moved from concept to clinical deployment in recent years. The first commercial product to implement the Pareto surface approach to MCO (from here onward, MCO will be taken to mean Pareto surface based MCO, which is the version of MCO studied in this report) is RayStation 2.0 (RaySearch Laboratories, Stockholm). In this implementation and in prototype implementations \cite{craftmonz, monz, thieke}, the Pareto surface plans, also called database plans, are fluence map optimized treatment plans. Fluence map plans are idealized plans: to make a plan clinically deliverable, the plan must be segmented into step and shoot segments or a dynamic sliding window plan. Segmentation causes the idealized fluence map plan to change, both because the fluence map ends up being approximated and because the dose calculation using multi-leaf collimator aperture shapes is a fundamentally different algorithm than the fluence map based dose calculation. 

This degradation of the treatment plan during segmentation can negatively impact the MCO treatment planning process by adding an iteration loop to the treatment planning process that the MCO technology was designed to avoid.  Anecdotally, this plan degradation happens more for complicated plans --  complicated here being synonymous with optimal isodose contours being highly non convex, i.e. the dose distribution being deeply carved away from OARs --  for example, bi-lateral target head and neck cancers and brain tumors situated near optic structures.

The purpose of this work is to describe a technique for IMRT Pareto surface plan navigation where the plans explored during navigation are already segmented, thus, ready-to-go.  In this sense, this work could also be called what you see is what you get (WYSIWYG) navigation. We assume step and shoot IMRT. The challenges of WYSIWYG navigation for step and shoot Pareto surface MCO can be understood by considering the following:

\begin{enumerate}
\item The Pareto surface should in general be expansive enough to cover a range of treatment planning possibilities. 
\item Plans far apart from one another on the Pareto surface will have different optimal segmentation, particularly when the number of segments is restricted.
\item A general point found during Pareto surface navigation is formed by a convex combination of up to $N$ of the pre-computed Pareto surface plans, where $N$ is the number of objective functions, i.e. the dimension of the Pareto surface. If each database Pareto optimal plan is pre-segmented into $S$ segments, then a navigated-to convex combination plan may have up to $N \times S$ segments.
\item It is highly desirable from a treatment delivery time perspective to have a plan which consists of small number of segments.
\end{enumerate}

\begin{figure}[htb]
\centering
\includegraphics[trim = 1mm 25mm 10mm 20mm, clip, width=13cm]{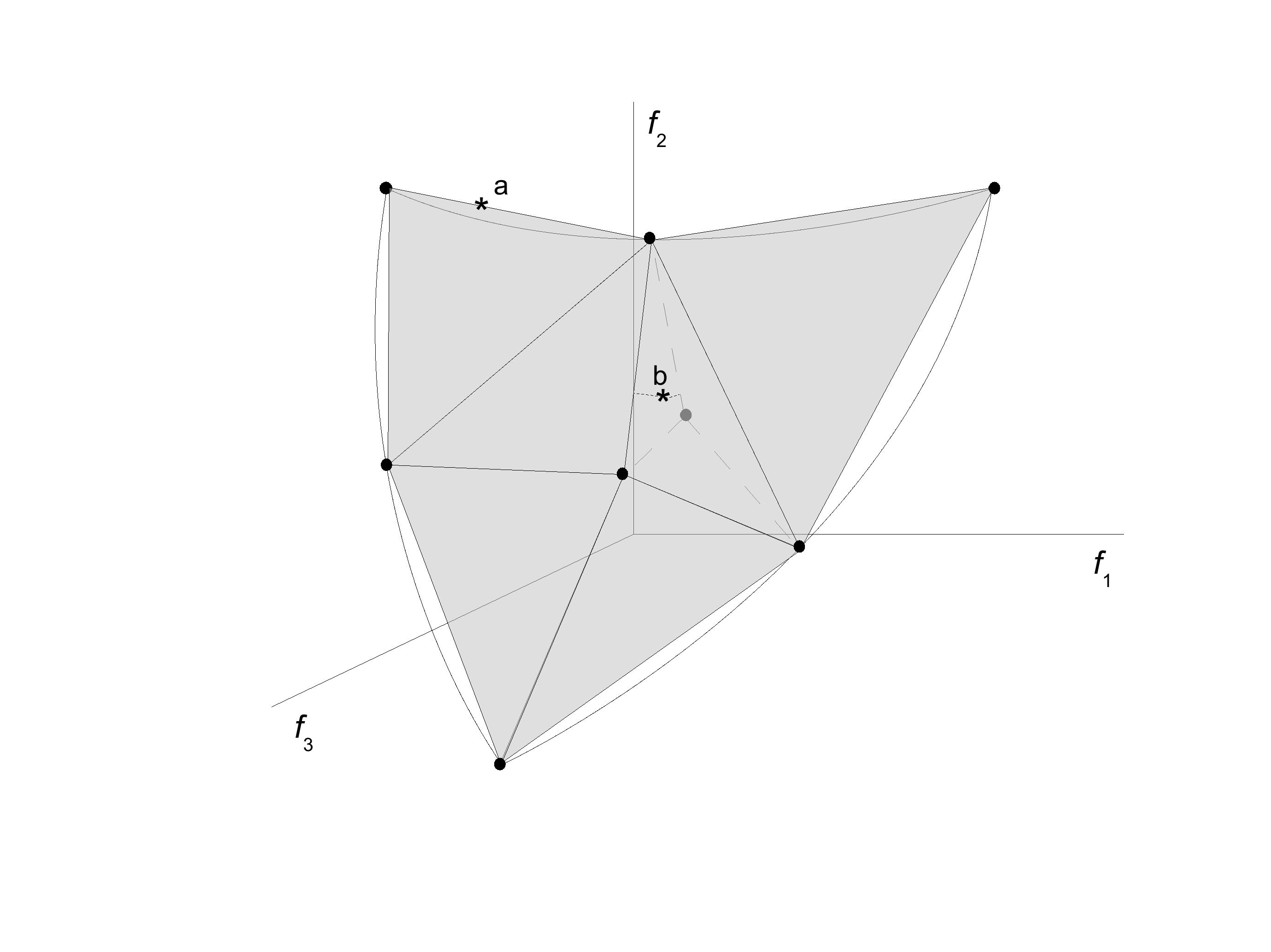}
\caption{\label{psurf} A 3D Pareto surface. The black circles indicate computed Pareto optimal plans, and the shaded triangles represent the convex combinations of those possible that give rise to undominated plans, and thus serve as an approximation of the Pareto surface. Points $a$ and  $b$ represent two convex combination plans that could occur during navigation. Since $a$ is on an edge, it is formed with fewer then $N=3$ plans. On the other hand, point  $b$ is in the center of the Pareto surface and thus requires $N$ plans to form it. However, depending on the number of plans used to represent the Pareto surface, it may be possible to reduce the number of plans needed to represent these central plans by snapping the convex combinations to nearby edge points, as shown by the two dotted line projections from point $b$. The shorter projection shows that if the number of computed plans on the surface is greater -- which we represent by adding another point, the gray point, to the surface and thus breaking that central facet into three smaller facets -- then the deviation will be less.}
\end{figure}

Together, these items point to the challenge. If we strive for a minimal set of segments for the final deliverable plan and we also strive to have a Pareto surface which encompasses an expansive tradeoff space, it is necessary to use a different segmentation for each of the database plans. 
During navigation it is theoretically possible to combine up to $N$ plans (in a clinical setting, $N$ can be 10 or more), depending on where you are on the Pareto surface. Plans along lower-dimensional edges of the Pareto surface use fewer than $N$ plans, see Figure~\ref{psurf} as well as section \ref{2dnav}. 


In this work we demonstrate that it is possible to continuously navigate a high dimensional discretely sampled Pareto surface while restricting the number of active plans used to represent the current location, without the restriction significantly impacting the plan quality. We demonstrate this in two ways. The first is a new navigation method based on visualizing 2D tradeoffs (cuts of the higher dimensional Pareto surface). We show analytically that this method naturally encourages the use of a small number of plans, and hence a small number of total segments, to form the navigation plans. The second is a statistical analysis of the more traditional ND navigation technique. For this we employ an automatic random navigation simulation and analyze the resulting statistics. Both approaches, one theoretical and one numerical, support the idea that the combination of only a few plans ($\ll N$) is sufficient to closely represent navigated positions on an $N$-dimensional Pareto surface, and this in turn supports the a concept of WYSIWYG step and shoot IMRT Pareto surface based navigation using pre-segmented database plans. 

\section{Methods}



We consider the following type of IMRT formulation:

\begin{eqnarray}
\hbox{optimize~~~} & \{f_1(d),~f_2(d), \ldots f_N(d)\} \nonumber\\
\hbox{subject~to~~~}   & d = Dx \nonumber\\
~~~~ & d \in C \nonumber\\
~ &  x \ge 0
\label{mco}
\end{eqnarray}

\noindent where $f_i$ are the $N$ objective functions describing the dose to a particular structure (e.g. mean dose to stomach), $d$ is the vector of voxel doses, $D$ is the dose-influence matrix, and $x$ is a concatenation of all the fluence maps into a single beamlet fluence vector. The set $C$ is a convex set of dose constraints (e.g. minimum dose to the target) that are designed to be met by all plans.\\

\subsection{Pareto surface generation and base plan segmentation}

For a given problem formulation, we use RayStation 2.0 to compute a set of points on the Pareto surface as follows. First the software computes the $N$ anchor plans. These are single objective optimizations of the individual objectives, i.e. the best possible plan for each of the objectives. The $N+1$th plan is termed the balance plan, and is found by optimizing an equi-weighted sum of the $N$ objectives. Objectives are similarly scaled so that an equi-weighted sum does indeed give a balanced plan.  For additional plans, RayStation optimizes equi-weighted sums of pairs of objective functions which are most highly anti-correlated. To determine level of correlation, the first $N+1$ plans are used to compute correlation coefficients between every pair of objectives. More advanced techniques for generating additional Pareto surface points, involving sandwiching the Pareto surface between lower and upper bounds and computing additional points to close the large gaps in these bounds, are described in \cite{bokrantz, craft-pgen, rennen} but are not used in this work. Here we compute a total of $2N$ plans using the anti-correlation method.
 

The key mathematical entity needed for navigation is the $M \times N$ matrix of objective function values, where $M$ is the number of pre-computed Pareto plans. In this report $M=2N$ as described above, but $M$ can be any number larger than $N$ in general. To access RayStation's $M \times N$ matrix for navigation simulation and evaluation outside of RayStation,  it needs to be extracted manually. This was done 
for the ``deliverable plan stage''. 
Thus, before navigation simulation, each base plan was segmented and for each plan, the values of the $N$ objective functions were read off, resulting in the $M \times N$ matrix.



With the Pareto matrix $P$ in hand, we analyze two styles of navigation. In both forms of navigation, the main idea is to form in real time convex combinations of the $M$ base plans (equivalently the $M$ rows of the Pareto surface matrix $P$) which respond to the planner's request to improve some objective. If the base plans are already segmented, such that all plans during navigation are ready-to-deliver, there is a strong delivery efficiency incentive to keep the number of base plans used in the convex combinations to a minimum.
The first navigation method we present and analyze theoretically, called {\em 2D-cut navigation}, is new to the field of Pareto surface exploration to our knowledge. It offers insight into why the number of active plans needed for convex combinations during navigation will often be small.  The next method, {\em ND navigation}, is the traditional $N$ sliders navigation method implemented in RayStation and prototype systems. 2D-cut navigation is presented because it offers a potentially more intuitive way to navigate high-dimensional surfaces, and also because one can analyze the plan-combining properties of it quite easily. However, given the dominance of ND navigation in the field, we numerically simulated the ND navigation style to show that it also has favorable plan-combining properties (low number of plans needed to approximate the user's navigation position on the Pareto surface).

\subsection{2D-cut navigation}
\label{2dnav}

The method described here is based on the display of a 2D cut of the Pareto surface. The user
selects which two objectives to view a 2D tradeoff curve of, and all of the other objectives are free, or 
upper bounded by a user-chosen value.  The user then navigates by either 1) moving along the 2D tradeoff curve
or 2) changing the bounds of one of the other objectives, which alters the 2D tradeoff curve when those constraints
affect the currently viewed tradeoff. At any point during navigation the user may switch which two objectives are currently 
active.


We assume that $P$ is ordered so that the anchor plans, in order, are the first rows of the matrix.
We normalize the Pareto surface via:
\begin{equation}
P_{k,j}^{\textrm{norm}}=\frac{P_{k,j}  - \min_m (P_{m,n=j})}    {\max_m (P_{m,n=j}) - \min_m (P_{m,n=j})}
\end{equation}
By expressing each objective in normalized form, between 0 and 1, the normalization assures equal influence of different objectives (or dimensions) for the multi dimensional distance calculation between different points on the Pareto surface that is introduced later. 
All calculations will be done in this normalized setting, and so without ambiguity we let $P$ from here onward refer to the normalized Pareto surface.


The Normal Constraint Method (NCM) can be used to generate an even distribution of points along the tradeoff curve for the two chosen objectives \cite{messac, messacBetter}.
Let $x$ and $y$ denote the two dimensions selected to view the tradeoff curve of (e.g. $x=7$ and $y=3$ would
be a possibility for say an 8D tradeoff).  Let $B$ denote the vector of constraints the user has selected.  This is 
the current position of the constraint sliders, and can include the objectives being traded off (in which case it is just the 
extent of the 2D curve that will be affected).

The NCM method starts by finding the anchor plans for the 2D tradeoff selected: 
\begin{eqnarray}
\hbox{minimize~~~} & (\lambda' P)_k \nonumber\\
\hbox{subject~to~~~}   & (\lambda' P)_j  \le B_j,~~~ \forall j \nonumber\\
 & \sum_{i=1}^{M} \lambda_i = 1 \nonumber \\
 & \lambda \ge0. 
\label{ancs}
\end{eqnarray}
$\lambda$ is the convex combination vector, and $k$ is the index of the objective to minimize (will run for $k=x$ and $y$).
This optimization finds the convex combination of plans in the database that minimizes the objective in question, subject to constraints $B$.

\begin{figure}[ht]
\centerline{\includegraphics[width=9cm]{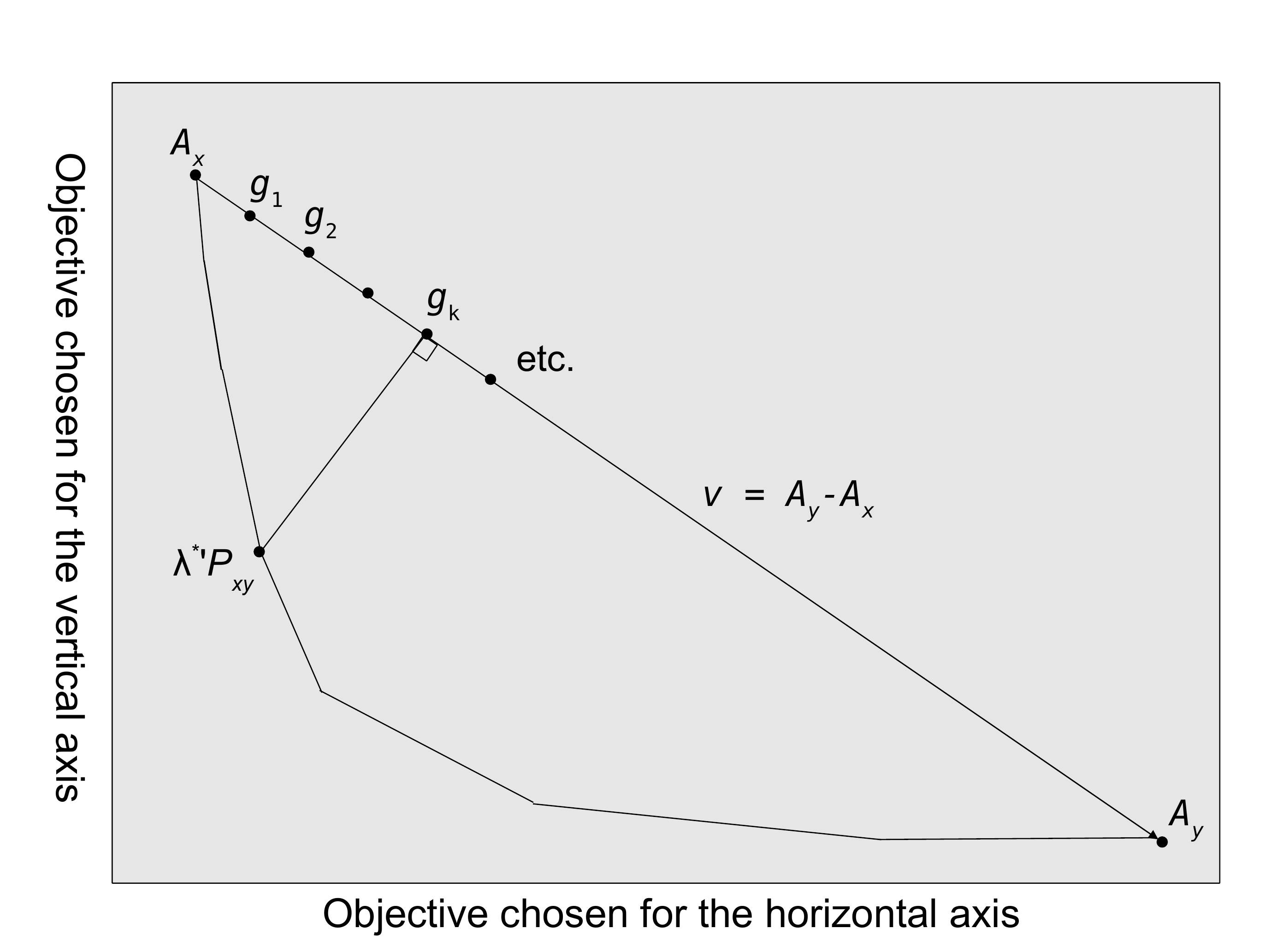}}
\caption{Diagram showing key idea of the normal constraint method (NCM) of computing points of a 2D tradeoff curve. $\lambda^{*'} P_{xy}$, the optimal solution of a single run of formulation \ref{nc}, is displayed.}
\label{2D}
\end{figure}

With those two anchors in hand, the next step is to form a moderate number of points (20 is likely to be more than enough) evenly spaced
along the line connecting them.  These are points in the 2D tradeoff space labeled $g_1,~g_2,\ldots,g_k,\ldots$ in Figure \ref{2D}.  From each of these points -- not including the anchor points -- we project down to find the intersection of the 2D tradeoff surface in the direction perpendicular to the line connecting the two anchor points.

To do this, we form the vector from anchor point $A_x$ to anchor point $A_y$, and call this $v$:
$$
v = A_y - A_x.
$$
$v$ is a $[2\times 1]$ vector.

Let $P_{xy}$ be the two user selected columns of the Pareto surface. Since there are $M$ plans in the Pareto database,
$P_{xy}$ is an $m \times 2$ matrix. Let $g_k$ be the point we are running. The following optimization yields the $k$th
point on the 2D Pareto curve:

\begin{eqnarray}
\hbox{minimize~~~} & (\lambda' P)_y \nonumber\\
\hbox{subject~to~~~}   & (\lambda' P)_j  \le B_j,~~~ \forall j \nonumber \\
 &  \lambda' P_{xy} v  =  g_k' v\nonumber \\
 & \sum_i \lambda_i = 1 \nonumber \\
 & \lambda \ge0. 
\label{nc}
\end{eqnarray}
The constraint $\lambda' P_{xy} v = g_k' v$ enforces that the solution is on the line perpendicular to the $v$ and passing through $g_k$.
Writing the constraint as $(\lambda'P_{xy} -g_k')v=0$ makes this easier to see. Note, the simpler method to generate points on the 2D curve would be to replace the $\lambda'(P_{xy} -g_k')v=0$ constraint with $ (\lambda' P)_x  \le b $ for various values of $b$ (the so-called epsilon constraint method of Pareto surface generation) but this does not have the useful property of spreading the
points evenly across the curve.  As for using a weighted sum method, we opt not for that because,
particularly for sparse Pareto surface representations (i.e. small number of database plans), the 2D
cuts will be very ``polyhedral'' (i.e. Pareto surface surface will consist of large facets) and so many weights will yield the same vertex.

Moving around on the 2D curve is then a matter of tracking which two points you are between and
updating the convex combination used. 

\subsection{On the number of active convex combination plans during 2D-cut navigation}
\label{proof}

\noindent Let $|B|$ denote the number of objectives that are constrained during 2D-cut navigation.

\noindent Proposition: The number of active plans needed to form the current navigation point is at most $2+|B|$.

\begin{proof}: Consider the linear programming formulation (\ref{nc}), which computes points on the 2D cut Pareto surface. Linear programming theory states that if a linear program is feasible and the optimal solution is bounded, there exists an optimal solution with DIM tight constraints (a constraint is tight if it is satisfied at equality), where DIM refers to the underlying dimension of the linear program, i.e. the number of optimization variables.  For linear program (\ref{nc}), there are $M$ optimization variables, the components of the  vector $\lambda$. Therefore, there exists an optimal solution to formulation (\ref{nc}) with $M$ tight constraints. Counting the number of constraints satisfied at feasible solution, we have the two equality constraints, we have (up to) $|B|$ navigation constraints. This means we need (at least) $M - (2+|B|)$ additional constraints to be tight. The only other constraints in this linear program are the $\lambda \ge 0$ constraints, so we need at least  $M - (2+|B|)$ components of $\lambda$ set to zero, or stated another way, at most  $2+|B|$ components of $\lambda$ positive.
\end{proof}

One caveat: linear interpolation between two of the points along the surface as computed by the NCM method could in theory double the number of underlying plans used. This is not in reality a problem since solving formulation \ref{nc} is very fast and therefore we can represent the 2D tradeoff curve easily with a high density of points and therefore avoid needing linear interpolation.


The above proof is sufficient to show that if the user invokes only a small number of constraints during navigation then the number of active plans used to form the current location on the Pareto surface is small.  But beyond this, we hypothesize that even more general Pareto surface navigation can also be done with maintaining a small set of active plans. To that end we turn to ND navigation.

\subsection{Traditional ND sliding navigation, and navigation simulation for generating statistics}\label{Automaticnavigation}
An automatic navigation approach was established to generate data to confirm the hypothesis that even in general navigation, the number of active convex combination plans needed is not large. 
In order to simulate navigation to a random spot on the Pareto surface, random constraint levels were chosen for each normalized objective based on a uniform probability density function. For each constraint level combination, points (plans) on the Pareto surface were calculated by minimizing each of the objectives individually and also by minimizing the equi-weighted sum of all objectives using optimization formulation \ref{ancs}. The tighter the randomly chosen constraint levels, the higher the probability that it is not possible to find any remaining plan on the Pareto surface. In those cases the constraint level combination is discarded, resulting in a modified distribution of the used constraint levels. 
These optimizations lead to a large set of $\lambda$ vectors, each representing a point of the Pareto surface. 

In a last step the random points on the Pareto surface $\lambda'P$ were approximated by plans with a restricted number of non-zero $\lambda_i$ values, so called ``small number of plans approximation'' resulting in $\tilde \lambda'P$. Two different methods have been used:
\begin{itemize}
\item Direct restriction of the number of active plans: Allow only a fixed number of $r$ non-zero plans. We chose the plans with the highest $\lambda_i$ values, set all other plan $\lambda_i=0$, and scale such that $\sum_{i=1}^{M}\tilde \lambda_{i}=1$.
\item Restriction of small $\lambda_i$ values: Set all $\lambda_i \leq \epsilon$ to zero and again scale remaining plans so that $\sum_{i=1}^{M}\tilde \lambda_{i}=1$.
\end{itemize}


To evaluate the difference between the original and the restricted plans, we use the average of the absolute objective function differences:

\begin{equation}
d(\lambda,\tilde \lambda)=\frac{1}{N} \sum_{j=1}^{N}\mid (\lambda'P)_{j}-(\tilde \lambda'P)_{j}\mid
\end{equation}

Plan quality after a $\tilde \lambda$ restriction is then defined as $Q(\lambda,\tilde \lambda)=1-d(\lambda,\tilde \lambda)$. Furthermore, the mean number of non-zero $\lambda_i$ and the distribution of $\lambda_i$-values were studied for every objective minimized.

\section{Results}

We analyze the following IMRT instance:

\begin{eqnarray}
\hbox{minimize~~~} &\{ \hbox{mean~left~eye~dose},~\hbox{mean~right~eye~dose},~\hbox{mean~chiasm~dose}, \nonumber\\
 \hbox{~~~~~~~~~~~~} & ~\hbox{mean~brainstem~dose},~\hbox{mean~left~cochlea~dose},~\hbox{mean~right~cochlea~dose}, ~~~~~~~~~~ \nonumber\\
 \hbox{~~~~~~~~~~~~} & ~\hbox{mean~right~optical~nerve~dose},~\hbox{mean~left~optical~nerve~dose}, ~~~~~~~~~~ \nonumber\\
 \hbox{~~~~~~~~~~~~} & ~\hbox{max~dose~to~PTV}, ~-\hbox{min~dose~to~PTV},~~~~~~~~~~ \nonumber\\
\hbox{~~~~~~~~~~~~} &  ~\hbox{dose~falloff~penalty:~59.4~Gy~to~0~Gy~in~1~cm} \}~~~~~~~~~~ \nonumber\\
\hbox{subject~to~~~}   & d = Dx \nonumber\\
~~~~ & d_i \ge 59.4 \hbox{~Gy},~ \forall i \in \hbox{~GTV}\nonumber\\
~~~~ & d_i \ge 40 \hbox{~Gy},~ \forall i \in \hbox{~PTV}\nonumber\\
~~~~ & d_i \le 70 \hbox{~Gy},~ \forall i \in \hbox{~skin}\nonumber\\
~~~~ & d_i \le 60 \hbox{~Gy},~ \forall i \in \hbox{~brainstem}\nonumber\\
~ &   x \ge 0
\label{mco-brain}
\end{eqnarray}

\begin{figure}
\centerline{\includegraphics[trim = 0mm 20mm 0mm 10mm, clip, width=16cm]{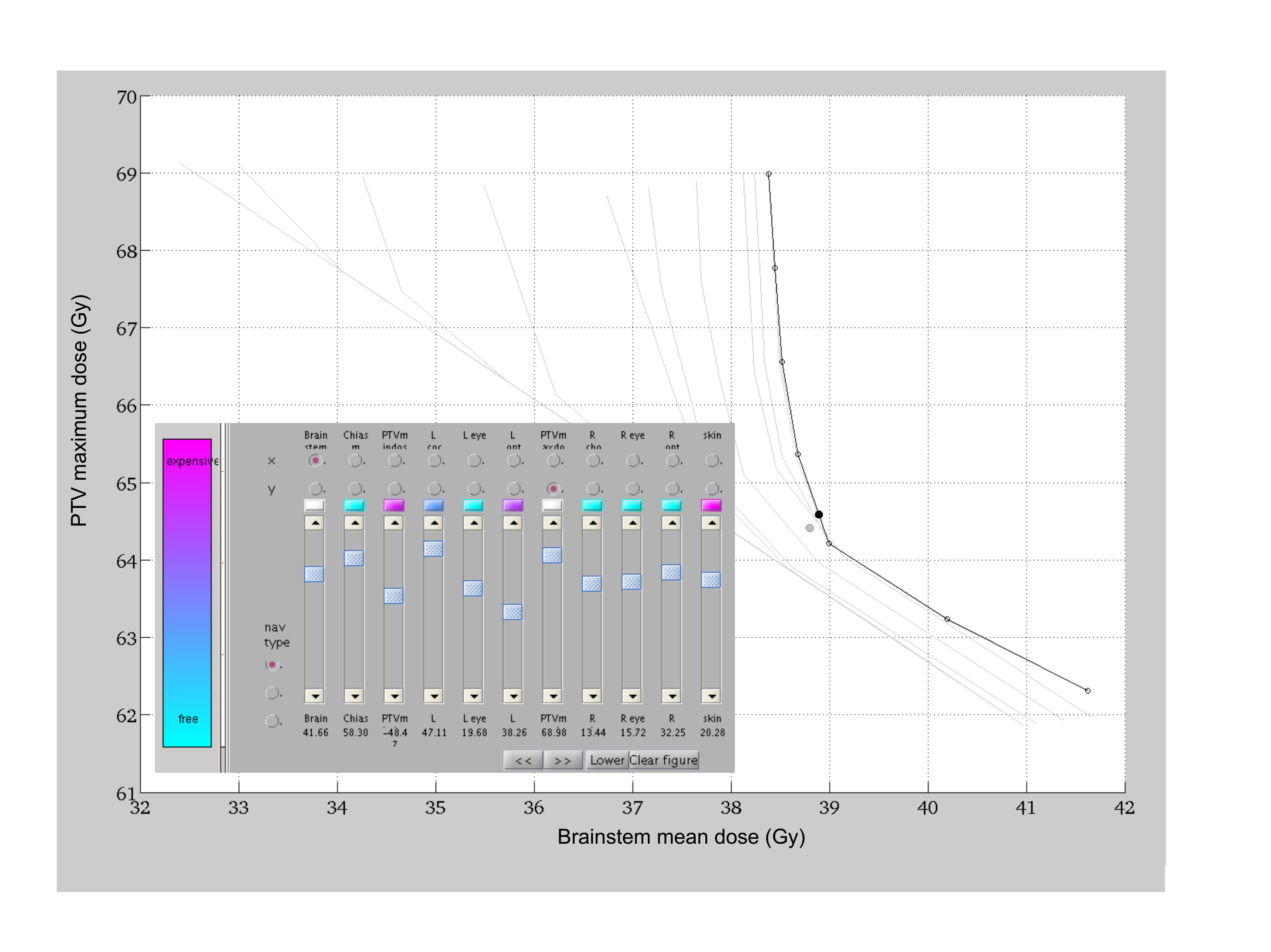}}
\caption{Here, the planner has chosen to view the tradeoff between brainstem mean dose and PTV maximum dose. The light gray Pareto curves display navigation history: the different curves arise from different constraint levels chosen for the other objectives. The black circle corresponds to the current navigation position without any $\lambda$ restrictions, and the gray circle below it corresponds to the plan after restricting $\lambda_i \ge 0.1$. The navigation interface is also displayed. The colors above each slider indicate the sensitivity of the Pareto curve to a change in that particular bound. The details of this computation are beyond the scope of this report, but briefly, linear programming duality theory is used to compute the sensitivity to changes in the constraint levels.}
\label{2dnavaction}
\end{figure}

The main result for the 2D cut navigation is the proof given in section \ref{proof}. However, to illustrate the 2D navigator and highlight the proof result, we display a snapshot of the navigation process in Figure \ref{2dnavaction}.
In this case we navigated to a plan that was a convex combination of 5 database plans. Using a cutoff value of 0.1, we reduce the number of active plans to 3, and the plan quality $Q = 99.2$\%. In this case, as viewed in the 2D plot in Figure \ref{2dnavaction}, it appears the restricted plan (gray circle) is actually better than the original plan (black circle) but this is only for the two chosen objectives: the plan will have degraded in some other objective(s).

\begin{figure}
\centerline{\includegraphics{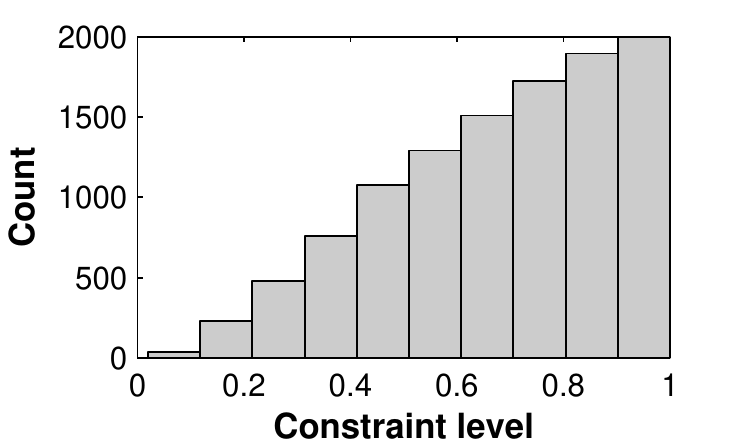}}
\caption{Distribution of the actual normalized constraint levels. 
  Starting with a uniform distribution of the normalized constraint levels, small constraint levels chosen simultaneously have a higher probability of an infeasible solution and are therefore more often discarded, leading to the triangle shaped distribution shown.}
\label{randnumb_distr}
\end{figure}

To study ND navigation, 12000 random points on the Pareto surface approximation were generated by minimizing each of the 11 objectives and a linear combination of all 11 objectives (1000 points for each objective minimized) as described in the Methods section. The number of segments $S$ for each database plans was chosen to be 50.  Figure \ref{randnumb_distr} shows the distribution of the (normalized) constraint levels that led to feasible plans. 
 Each of the 12000 randomly chosen Pareto combinations was approximated with both methods described in section \ref{Automaticnavigation}. 

\begin{figure}
\begin{centering}
\textbf{Unrestricted}\vspace{-3mm}
\par\end{centering}{\large \par}
\begin{centering}
\subfloat{\includegraphics[viewport=0bp 0bp 227bp 123bp,clip]{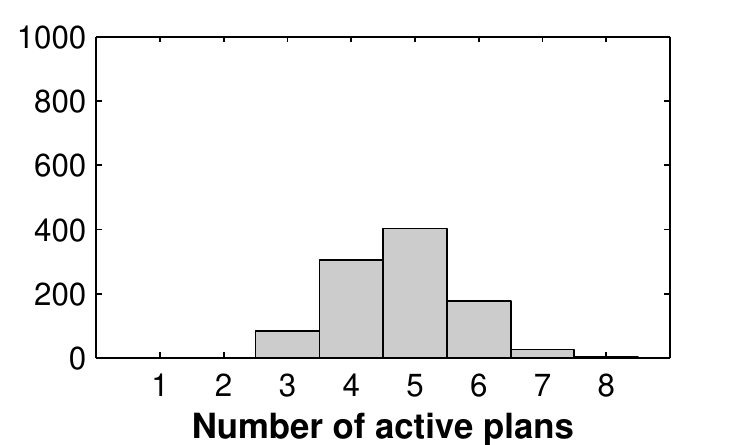}}\hspace{2mm}\subfloat{\includegraphics[viewport=0bp 0bp 227bp 123bp,clip]{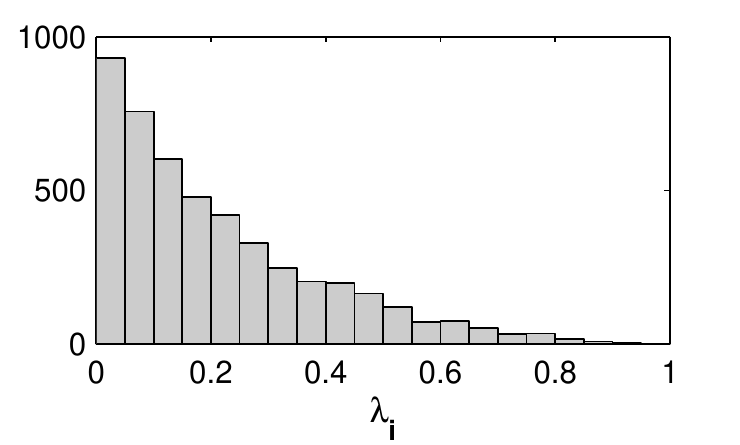}}\vspace{2mm}
\par\end{centering}
\begin{centering}
\textbf{Restrict number of active plans: $n_{\lambda_{i} \neq 0}\leq3$}\vspace{-4mm}
\par\end{centering}{\large \par}
\begin{centering}
\subfloat{\includegraphics[viewport=0bp 0bp 227bp 123bp,clip]{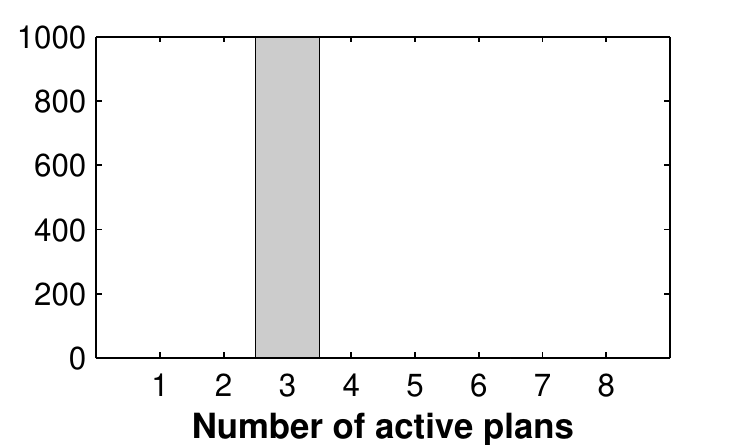}}\hspace{2mm}\subfloat{\includegraphics[viewport=0bp 0bp 227bp 123bp,clip]{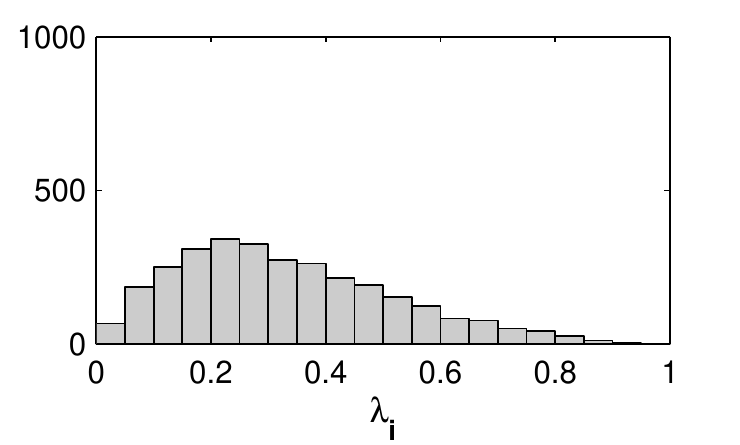}}\vspace{2mm}
\par\end{centering}
\begin{centering}
\textbf{Restrict values of $\lambda_{i}$: $\lambda_{i}\geq0.1$}\vspace{-4mm}
\par\end{centering}{\large \par}
\begin{centering}
\subfloat{\includegraphics[viewport=0bp 0bp 227bp 123bp,clip]{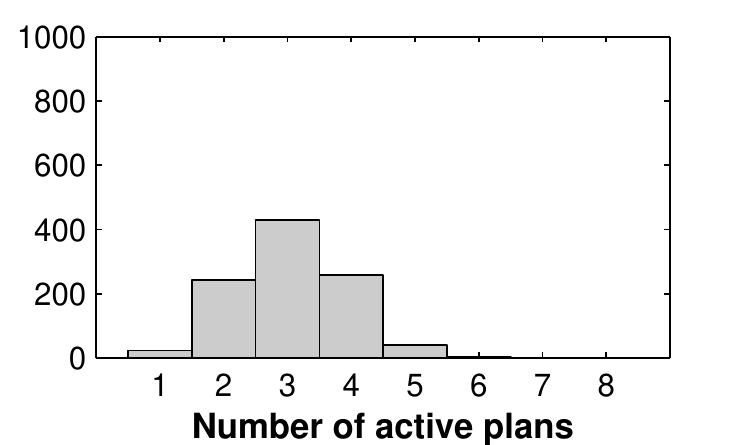}}\hspace{2mm}\subfloat{\includegraphics[viewport=0bp 0bp 227bp 123bp,clip]{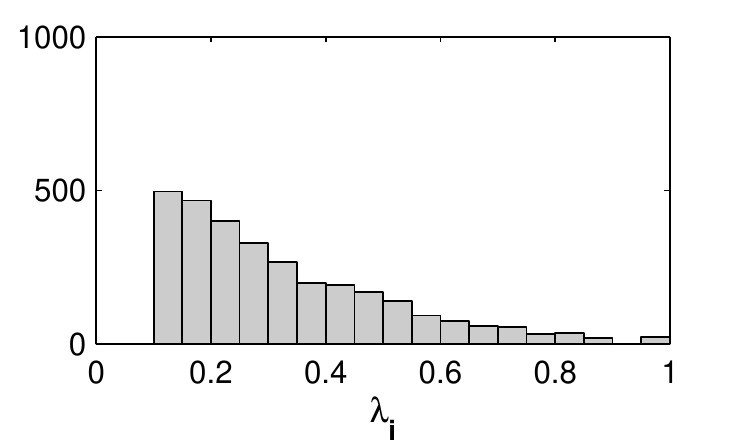}}
\par\end{centering}
\caption{Histograms of the number of active plans and also of the $\lambda_i$ values for the calculated 1000 Pareto surface points by minimizing the {\bf PTV-minimal dose objective}. Three different cases are shown: (a) unrestricted plans, (b) restriction to a total number of 3 or fewer active plans and (c) restriction to $\lambda_i \geq 0.1$.}
\label{statistics_for_obj3}
\end{figure}

\begin{figure}
\begin{centering}
\textbf{Unrestricted}\vspace{-3mm}
\par\end{centering}{\large \par}
\begin{centering}
\subfloat{\includegraphics[viewport=0bp 0bp 227bp 123bp,clip]{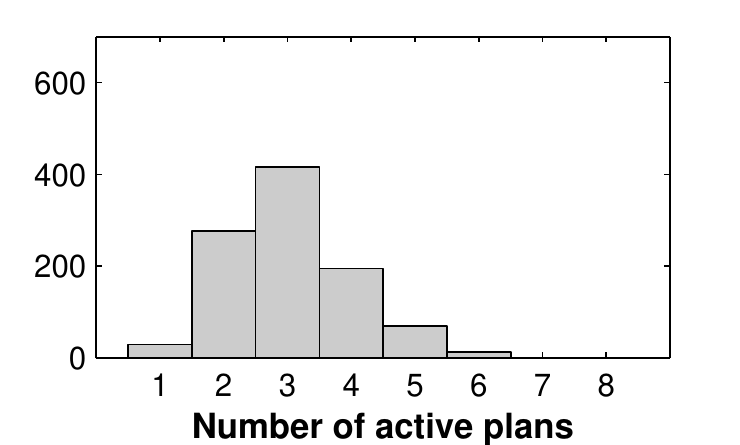}}\hspace{2mm}\subfloat{\includegraphics[viewport=0bp 0bp 227bp 123bp,clip]{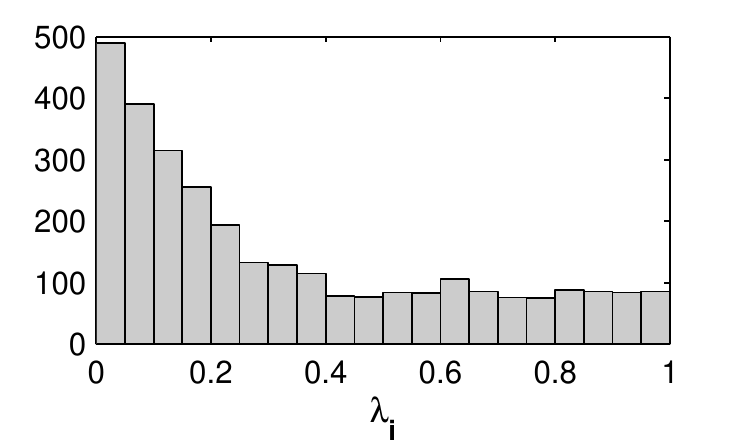}}\vspace{2mm}
\par\end{centering}
\begin{centering}
\textbf{Restrict number of active plans: $n_{\lambda_{i} \neq 0}\leq3$}\vspace{-4mm}
\par\end{centering}{\large \par}
\begin{centering}
\subfloat{\includegraphics[viewport=0bp 0bp 227bp 123bp,clip]{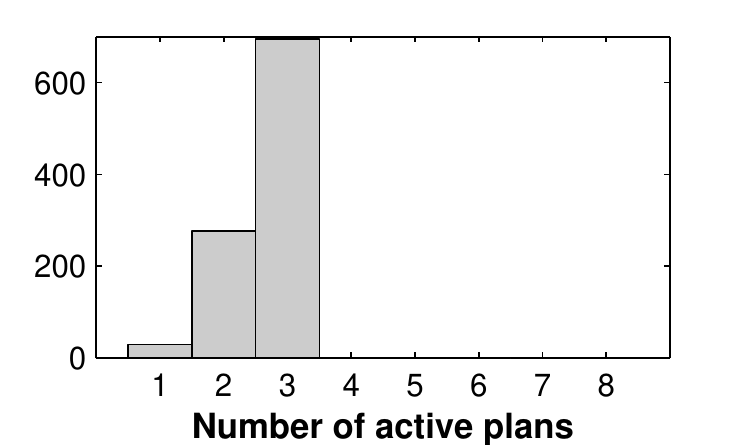}}\hspace{2mm}\subfloat{\includegraphics[viewport=0bp 0bp 227bp 123bp,clip]{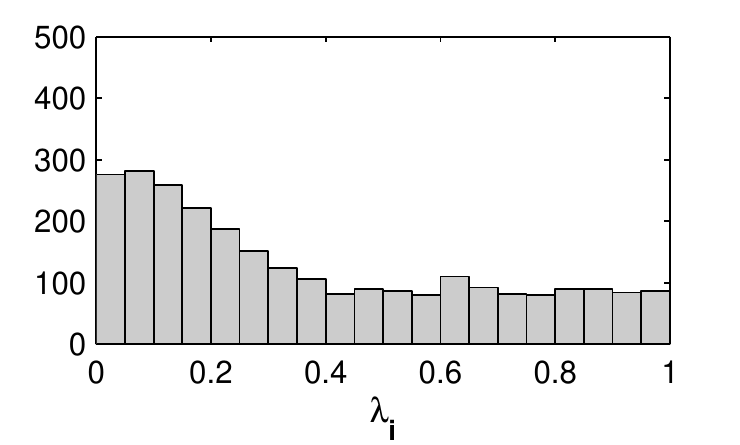}}\vspace{2mm}
\par\end{centering}
\begin{centering}
\textbf{Restrict values of $\lambda_{i}$: $\lambda_{i}\geq0.1$}\vspace{-4mm}
\par\end{centering}{\large \par}
\begin{centering}
\subfloat{\includegraphics[viewport=0bp 0bp 227bp 123bp,clip]{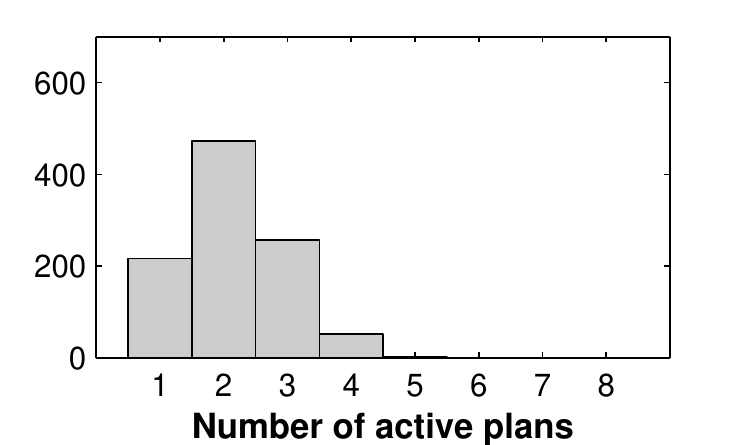}}\hspace{2mm}\subfloat{\includegraphics[viewport=0bp 0bp 227bp 123bp,clip]{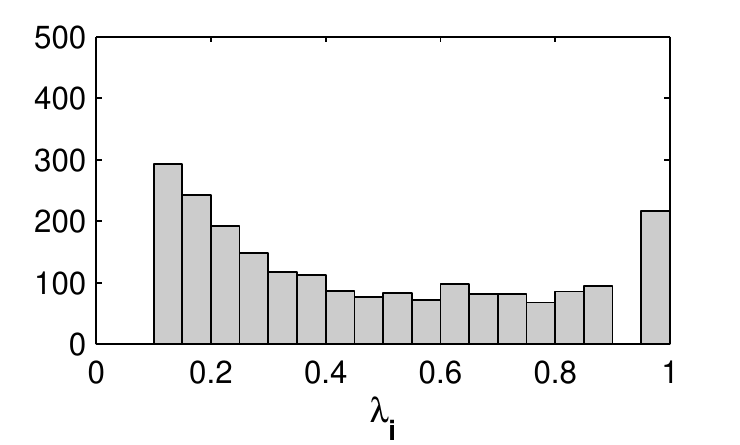}}
\par\end{centering}
\caption{Histograms of the number of active plans and also of the $\lambda_i$ values for the calculated 1000 Pareto surface points by minimizing the {\bf Linear combination objective}. Three different cases are shown: (a) unrestricted plans, (b) restriction to a total number of 3 or fewer active plans and (c) restriction to $\lambda_i \geq 0.1$.}
\label{statistics_for_obj12}
\end{figure}
Figures \ref{statistics_for_obj3} show histograms of the number of active plans and also of the $\lambda_i$ values for the calculated 1000 Pareto surface points from minimizing the (negative) PTV-minimal-dose objective. 
Three different cases are shown: (a)  before restriction, (b) after restriction to a total number of 3 or fewer active plans and (c) after restriction to $\lambda_i \geq 0.1$. The chosen PTV-minimal-dose objective was the hardest to be fulfilled, meaning that the mean number of plans used was the greatest compared to the other objectives. In Figure  \ref{statistics_for_obj12} the same analysis is shown for the 1000 points on the Pareto surface obtained when minimizing the linear combination of all 11 objectives. The linear combination objective required the fewest active $\lambda_i$ on average.
\begin{table}
\setlength{\extrarowheight}{4pt}
\centering{}%
\begin{tabular}{>{\raggedright}m{3cm}>{\raggedright}m{4.7cm}>{\centering}m{3.5cm}>{\centering}m{3.5cm}}
\hline 
\multirow{2}{3cm}{} & \multirow{2}{2.5cm}{} & \multicolumn{2}{c}{\textbf{\small Objective minimized}}\tabularnewline
 &  & \textbf{\small PTV minimal dose} & \textbf{\small Linear combination}\tabularnewline
\cline{1-4} 
\multirow{3}{3cm}{\textbf{\small Mean number of active plans}} & {\small unrestricted} & {\small $4.77\pm0.95$} & {\small $3.04\pm1.00$}\tabularnewline
 & {\small restricted to $n_{\lambda_{i} \neq 0}\leq3$} & {\small $3.00\pm0.00$} & {\small $2.67\pm0.53$}\tabularnewline
 & {\small restricted to $\lambda_{i}\geq0.1$} & {\small $3.06\pm0.89$} & {\small $2.15\pm0.82$}\tabularnewline
\cline{1-4} 
\multirow{2}{3cm}{\textbf{\small Plan quality $Q$}} & {\small restriction to $n_{\lambda_{i} \neq 0}\leq3$} & {\small $0.973\pm0.022$} & {\small $0.994\pm0.013$}\tabularnewline
 & {\small restriction to $\lambda_{i}\geq0.1$} & {\small $0.978\pm0.019$} & {\small $0.984\pm0.019$}\tabularnewline
\hline 
\end{tabular}\caption{Statistics of the number of active plans in the unrestricted case and after applying both restriction methods. The plan quality $Q$ for two examples of both restriction methods is shown. Given uncertainties represent the $1\sigma$ confidence interval of the variation over the 1000 single Pareto surface points.}
\label{tab_statistics}
\end{table}
Table \ref{tab_statistics} summarizes the mean number of active plans and the plan quality for both, the PTV-minimal-dose objective and the linear combination objective.

\begin{figure}[ht]
\subfloat[]{\includegraphics{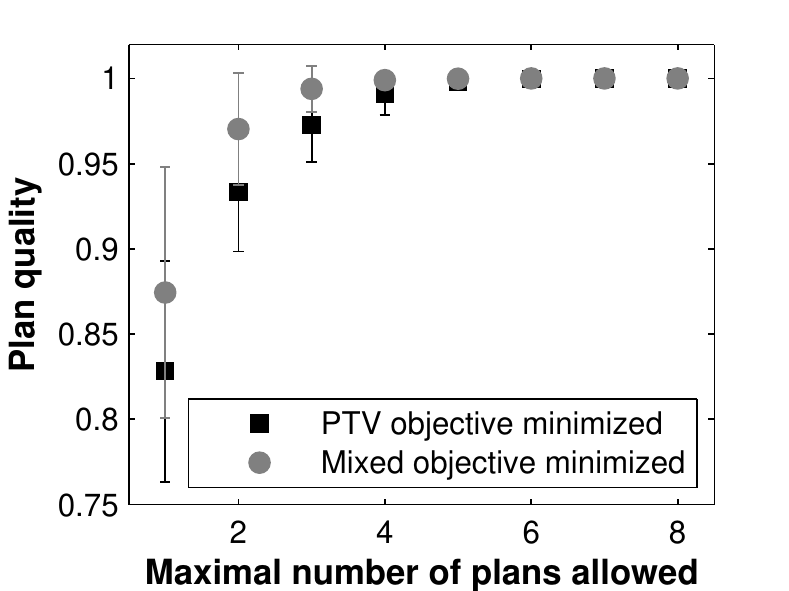}}
\hspace{5mm}
\subfloat[]{\includegraphics{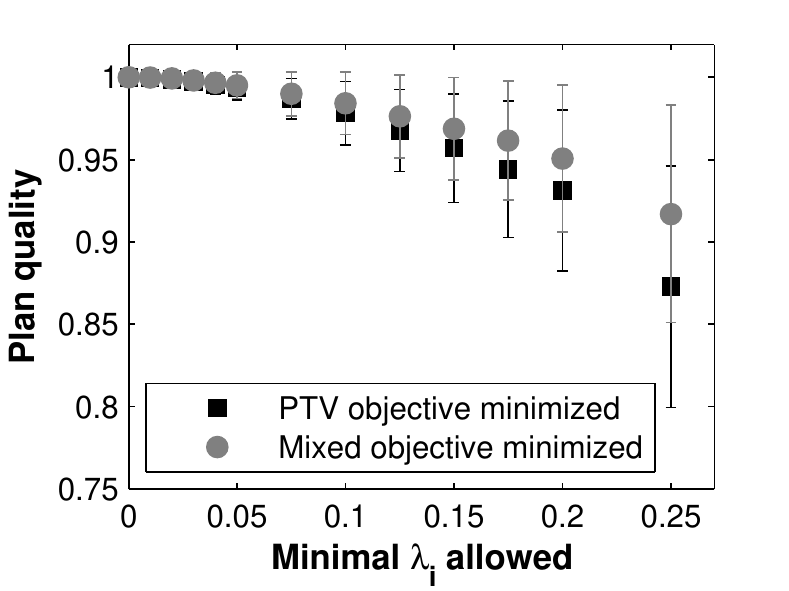}}
\caption{Plan quality for different restriction levels for both restriction methods applied: (a) number of active plans restriction and (b) $\lambda_i$ value restriction. Data points are shown for two selected objectives minimized. Each data point corresponds to the mean value of 1000 calculated Pareto surface points. Uncertainty bars represent the standard deviation.}
\label{PQ}
\end{figure}

Comparing both cases it can be seen that the $\lambda_i$ distribution is tilted more to the left side  with nearly no $\lambda_i \geq 0.9$ for the hard case compared to the easy case. Comparing the $\lambda_i$ distribution for both restriction methods, it becomes clear that the restriction of the maximal number of plans still leads to a significant amount of very small $\lambda_i$ values although their number is reduced compared to the unrestricted case. Obviously, this is not the case with the second restriction method. The fact that both restriction methods lead to approximately the same mean number of active plans is not an intrinsic characteristic of the restriction method, but only caused by the chosen restriction levels for both methods.

In fact, both restriction methods have been applied for different restriction levels, meaning (a) different numbers of active plans allowed or (b) different minimal values of $\lambda_i$. For both cases Figure \ref{PQ} shows the plan quality for different restriction levels and the two selected objectives from above. For the same restriction applied, the approximation of the Pareto surface points generated by minimizing the linear combination objective lead to better plan quality than the approximation of the Pareto surface points generated by minimization the PTV dose objective. Restricting the Pareto surface points to a high number of plans ($\geq 5$) or to a low minimal $\lambda_i$ value ($\lambda_i \leq 0.05$) leads to nearly no degradation of the plan quality.

\begin{figure}[ht]
\centerline{\includegraphics{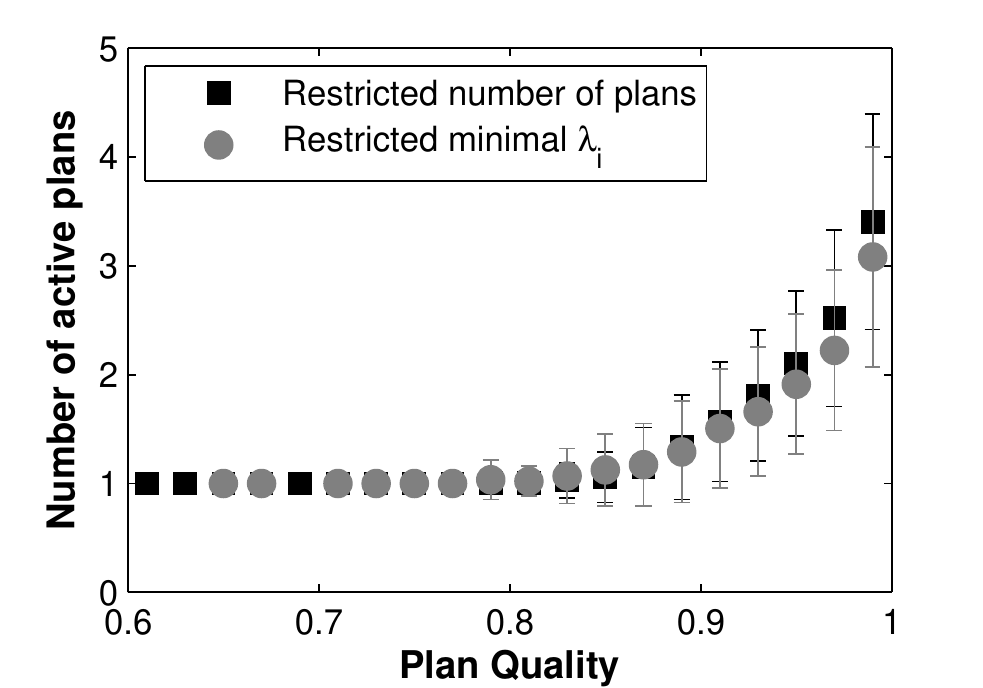}}
\caption{Number of active plans plotted as a function of plan quality for both restriction methods. The plot contains all different restrictions shown in the previous figure for all 11+1 objectives minimized. The resulting plan quality was binned for each of both restrictions methods into bins with a bin size of 0.02. Therefore each data point corresponds to all restrictions resulting in a plan quality of $\pm 0.01$ from the x-coordinate of the data point. Uncertainty bars represent the standard deviation.}
\label{fig_numberofplans_over_pq}
\end{figure}

To evaluate how many plans are actually active for a given plan quality to be fulfilled, in Figure \ref{fig_numberofplans_over_pq} the number of active plans is plotted over the plan quality achieved. The plot includes the data of all different restriction levels shown in the previous Figure to all 12000 Pareto surface points. From the presented data it can be seen that for a given plan quality on average fewer plans are used with the restriction of the $\lambda$ values compared to the restriction of the maximal number of plans. This finding is in agreement with our understanding of the different restriction methods. The restriction to a maximal number of plans eliminates plans if the total number of non-zero plans is higher than allowed -- even if they have a high contribution (a high $\lambda_i$). On the other hand, it does not remove plans with small contributions if the total number of non-zero plans is equal or below the allowed number. Both this is not the case for the second method -- the restriction in the $\lambda$ value space. Moreover this method is more flexible when going to different cases with different number of objectives. Then the same $\lambda_i$-restriction can still be applied, but a restriction to a maximal number of plans obviously needs to be adjusted. Therefore, we conclude that restricting the $\lambda$-space is the method of choice for reducing the number of plans.

\begin{figure}[ht]
\includegraphics[width=16.5cm]{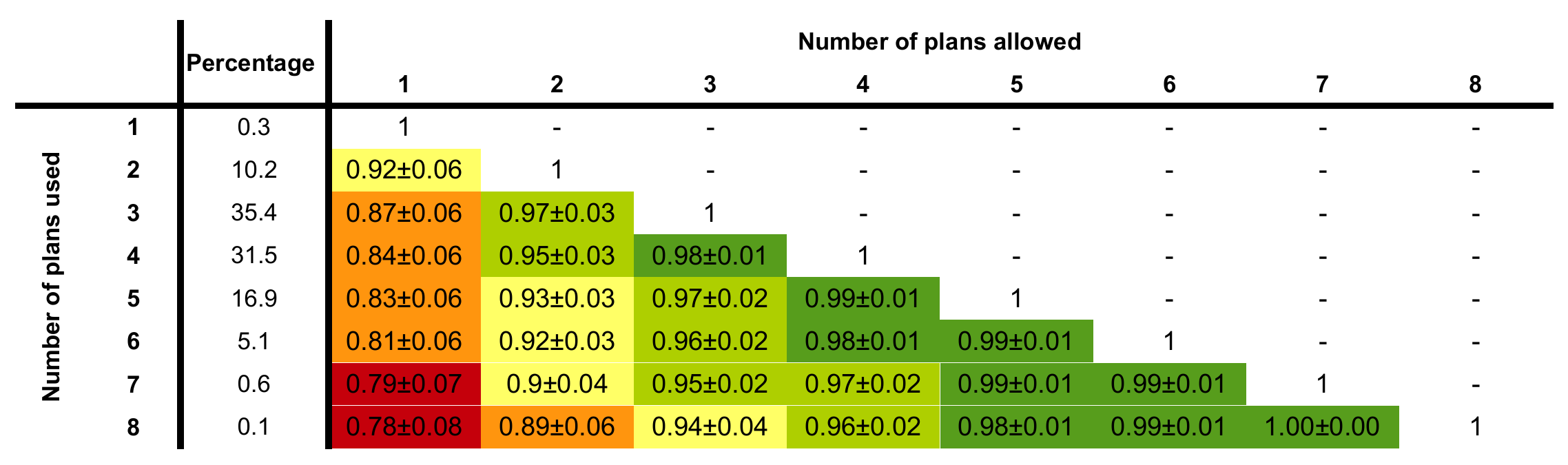}\caption{Plan quality matrix: Plan quality degrades as you allow fewer plans for plan representation. Number of plans used refers to the unrestricted number of plans during standard navigation. When this number is restricted to a smaller number, shown across the top of the matrix, the plan quality drops. Still, restricting to 3 plans retains a plan quality of over 95\% except in the rare instance, 1 out of 1000 in this case, where 8 plans are used in the unrestricted setting.} 
\label{pq_table}
\end{figure}

Figure \ref{pq_table} shows the plan quality for all plans with a specific number of plans before and after application of the restriction. It can be seen that in general three plans are enough to achieve a plan quality of at least 0.95. Only in the vary rare cases ($0.1\, \%$ of the initial random plans) when there are more than 7  plans used in the initial plan, the plan quality will  be reduced by more than $5\, \%$ on average. 
An interesting result is also that already $46\, \%$  of the initial random plans use no more than 3 plans.

\begin{figure}[ht]
\includegraphics[width=16.5cm]{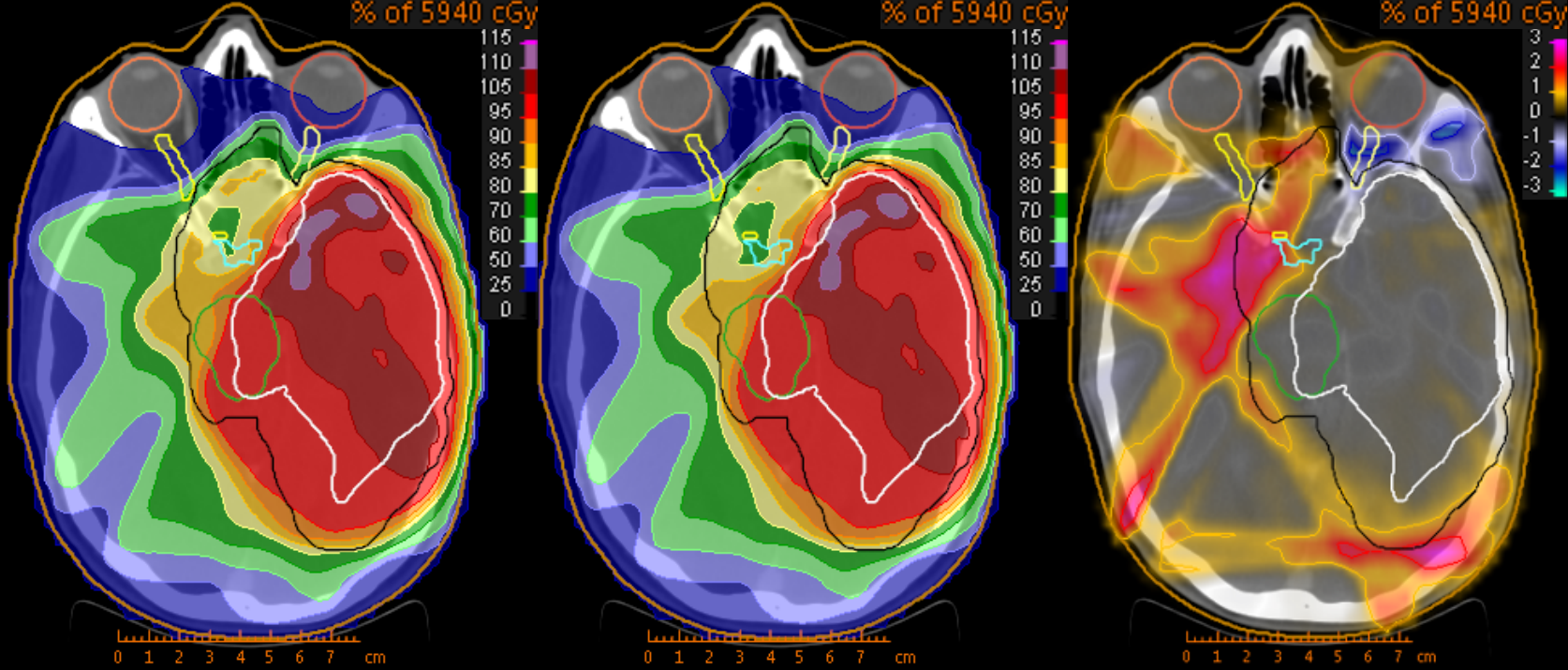}\caption{Dose wash of a random plan with 6 plans before (left) and after restricting the number of plans to 3 (middle). On the right the dose difference is shown. Contours mark PTV (black), GTV (white) and organs at risk (green, light blue, dark and light yellow, dark and light orange).}
\label{dosewash}
\end{figure}
\begin{figure}[ht]
\includegraphics[width=16.5cm]{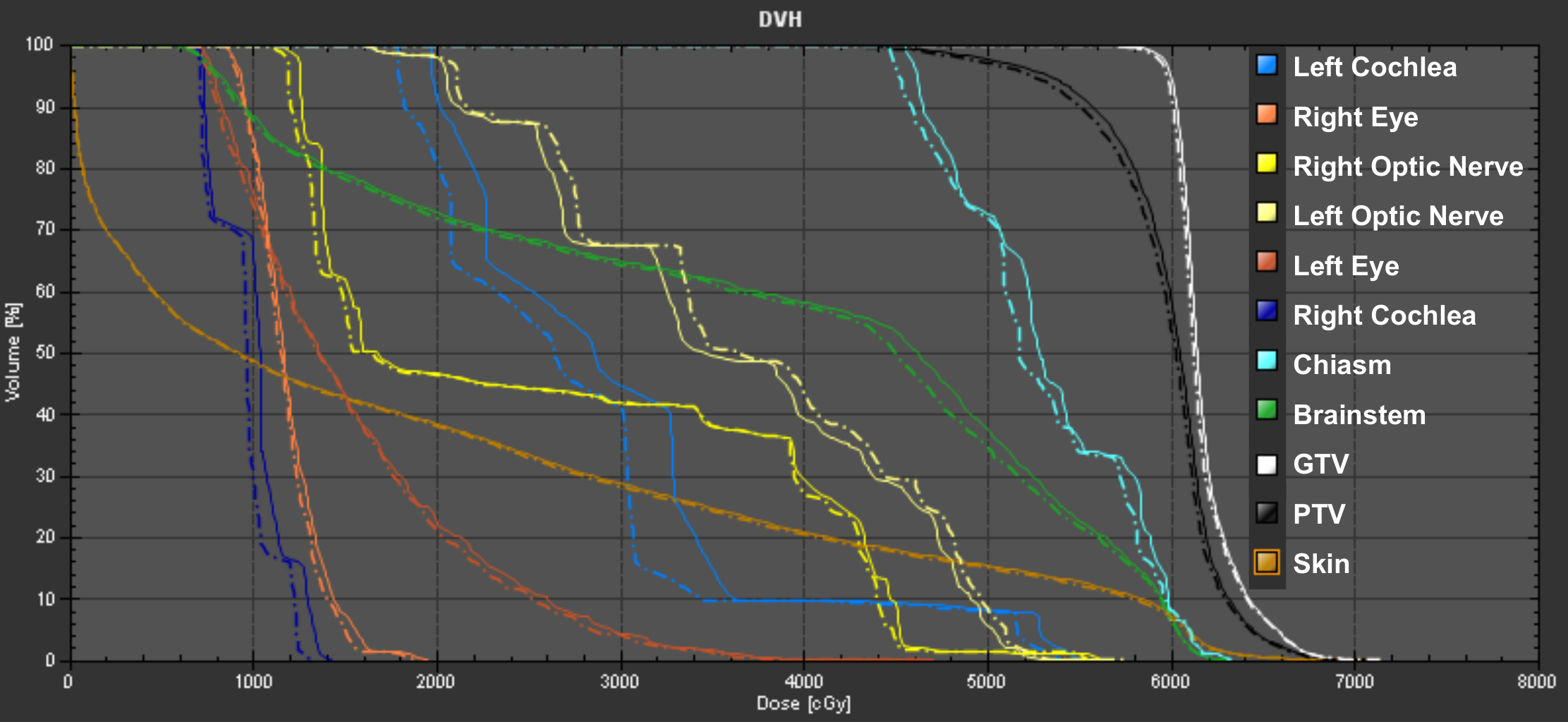}\caption{Dose volume histogram for the case shown in Figure \ref{dosewash}. The continuous and dashed lines represent different structures before and after the application of the restriction, respectively.}
\label{dvh}
\end{figure}

In Figure \ref{dosewash} the dose distribution of a random case that uses 6 plans and that was restricted to 3 plans is shown. For selection of this case, we ordered all cases that used 6 plans before and 3 plans after the application of the restriction according to their plan quality. The plan corresponding to the median plan quality of these cases is the one presented. After restriction, this plan showed a plan quality of $96.1 \, \%$. The dose difference after restriction is below $4\, \%$ in all voxels. The dose volume histograms (DVH) before and after restriction are shown in Figure \ref{dvh}, which confirms that only minor deviations occur as a result of the restriction. 

\section{Discussion and conclusions}
This work grew out of the desire to avoid the ``create deliverable plan'' step in the current MCO navigation technique. Imagining what an ideal system would look like, where one is sliding on a Pareto surface and observing dose volume histograms and isodose contours update continuously and in real time, it becomes apparent that the MLC apertures used should depend on where one is on the Pareto surface. Initial considerations of this had individual MLC segment shapes coming in and out of the active set of apertures. Using this technique however would require a much larger real time computation, since one would need to maintain in memory the dose to voxel contributions of the individual apertures and make sure the composite dose from the current set of apertures is feasible and optimal. Instead, we have relied on the following observation in this work: {\em a particular aperture makes sense in conjunction with the other apertures it was created with.} Stated another way, a particular aperture is useful because it is part of a full collection of apertures which together make a good treatment plan. Therefore, we can simplify the navigation considerably by swapping apertures in and out not individually, but collectively as the entire plan that they were created for, and scaled as a group. This allows us to consider simply the dose cubes (the term dose cube means the entire dose distribution), or even more minimally, the functions of the dose cubes that serve as the plan descriptors (e.g. minimum target dose, mean brainstem dose) during navigation. 

In order for this idea of navigating by mixing pre-segmented plans to be feasible, 
it needs to be the case that not too many plans need to be combined during navigation. 
In this work we have demonstrated that a limited number of plans suffices to approximate 
a general location on a high dimensional Pareto surface. We demonstrate for a brain case with an 11 dimensional Pareto surface, generally 5 or fewer plans are needed to form the navigated plan, and that even for plans requiring 6 base plans, 3 plans is enough for very close approximation, see Figure \ref{pq_table}. We also show that instead of requiring a combinatorial search for ``what 3 plans'' to use, it is sufficient to use the plans of the original convex combination that have the highest contributions (the highest $\lambda_i$ values). Assuming each pre-computed Pareto surface plan is segmented into say 50 segments, as was done in this work, the total number of segments will be $3\times50 = 150$ or fewer (fewer if the same segments were used in more than one of the plans).

A problem arises when one considers that if 50 plans is enough to generate a good plan, than 150 plans is overkill and therefore not a desirable solution. On the other hand if around 150 segments are needed, then the base plans, with only 50 segments, might be far from Pareto optimal. We next present two different lines of attack to deal with this issue.  

As the number of plans on the original surface is increased, the error in forming a navigated plan with a restricted number of active plans will decrease. Indeed in the limit of a very fine grid of plans pre-computed on the surface, one could use a single plan -- the closest one -- to represent a given convex combination plan, see Figure \ref{psurf}. Populating a Pareto surface with a fine grid can be done in parallel and offline, and real-time navigating a surface with hundreds of plans is possible due to the simplicity of the navigation optimization problems that are solved, which do not involve the dose cube or the beamlets. In this mode of operation one would segment each database plan with as many segments are needed or clinically allowed for delivery time considerations and the navigation system would display the plan with the highest $\lambda_i$ value.

If the number of pre-computed plans cannot be made large enough to allow for single plan navigation, or mixing is desired for smooth exploration of the planning options, then shared segments can also alleviate the problem of the pre-computed plans needing to be too coarsely segmented and thus not Pareto optimal. In this case one would find a pool of apertures that are used by all the plans and then additional specialized apertures for each plan. Direct aperture optimization (e.g. \cite{dao,dmpo, dao2}) will be immediately applicable here. 

The approach presented is a practical compromise, conceived to allow smooth MCO exploration of step and shoot plans without a large computational overhead. Ideally as the planner navigates the Pareto surface the number of segments would be changing to reflect the varying complexity of plans in different regions of the Pareto surface. Additionally, one would like to be able to explore the benefit of adding a few more apertures during navigation. In the current scheme, apertures come as a full plan bundle, so these features are not handled smoothly. Despite the drawbacks, we see the presented approach as a practical alternative to real time aperture optimization while sliding over the Pareto surface, which at the current state is not computationally feasible. It should also serve as groundwork for future developments on directly deliverable step and shoot MCO, which in its more mature state will likely employ a hybrid approach of a pool of apertures shared by all of the database plans, specialized apertures for different plans on this Pareto surface, limiting the number of plans combined during navigation, and possibly real-time aperture weight optimization and the fast creation of helpful additional apertures. Future research will also be directed towards directly deliverable navigation for sliding window IMRT planning, which will require a different line of attack.

In terms of why so few plans are needed during Pareto navigation, we have provided an explanation in reference
to the newly introduced 2D-cut navigation, where we show by linear programming theory that the number of active variables in an optimal solution is related to the number of function constraints that are binding during navigation. Another view of why the number of active plans will typically be much less than the theoretical maximum number of $N$ is related to the fact that for high dimensional Pareto surfaces in IMRT planning, many of the objectives will be strongly correlated, leading to effectively lower dimensional Pareto surfaces, i.e. an effectively smaller $N$. For example, in a brain case, one might expect that minimizing the dose to the left eye also minimizes the dose to the left optic nerve.  The high dimensional surfaces analyzed by Spalke {\em et al} \cite{spalke} were shown to have an effective dimension of 3 or 4 when the dimensions were reduced using principal components analysis or a non-linear technique called the isomap method \cite{isomap}. 

Both 2D and ND navigation analysis lead us to the same conclusion: relatively few plans are needed to approximate an arbitrary navigated-to location on an IMRT Pareto surface. This finding justifies our proposal for direcly deliverable step and shoot IMRT MCO: pre-segment the base Pareto surface plans and restrict the number of plans used during navigation to form the convex combination plans. Direct deliverability is a crucial step to bringing the full power of MCO into clinical IMRT treatment planning.

\section*{Acknowledgements}
This work was supported by NCI Grant 1 R01 CA103904-01A1: Multi-criteria IMRT Optimization and by a research
collaboration with RaySearch Laboratories. Thanks to Ehsan Salari for helpful feedback on the manuscript.
\clearpage

\bigskip
\bibliographystyle{unsrt}
\bibliography{all}

\end{document}